\pdfoutput=1
\documentclass[preprint,prb]{revtex4}
\usepackage{bm}
\usepackage{color}
\usepackage{graphics}
\usepackage{setspace}
\usepackage{array}
\usepackage{amsmath}

\begin{document}
\title{Competing quantum effects in the dynamics of a flexible water model}
\author{Scott Habershon, Thomas E. Markland and David E. Manolopoulos}
\affiliation{Physical and Theoretical Chemistry Laboratory, University of Oxford, South Parks Road, Oxford, OX1 3QZ, UK}

\begin{abstract}
Numerous studies have identified large quantum mechanical effects in the dynamics of liquid water. In this paper, we suggest that these effects may have been overestimated due to the use of rigid water models and flexible models in which the intramolecular interactions were described using simple harmonic functions. To demonstrate this, we introduce a new simple point charge model for liquid water, q-TIP4P/F, in which the O--H stretches are described by Morse-type functions. We have parameterized this model to give the correct liquid structure, diffusion coefficient, and infra-red absorption frequencies in quantum (path integral-based) simulations. The model also reproduces the experimental temperature-variation of the liquid density and affords reasonable agreement with the experimental melting temperature of hexagonal ice at atmospheric pressure. By comparing classical and quantum simulations of the liquid, we find that quantum mechanical fluctuations increase the rates of translational diffusion and orientational relaxation in our model by a factor of around 1.15. This effect is much smaller than that observed in all previous simulations of simple empirical water models, which have found a quantum effect of at least 1.4 regardless of the quantum simulation method or the water model employed. The small quantum effect in our model is a result of two competing phenomena. Intermolecular zero point energy and tunneling effects destabilize the hydrogen bonding network, leading to a less viscous liquid with a larger diffusion coefficient. However this is offset by intramolecular zero point motion, which changes the average water monomer geometry resulting in a larger dipole moment, stronger intermolecular interactions, and slower diffusion. We end by suggesting, on the basis of simulations of other potential energy models, that the small quantum effect we find in the diffusion coefficient is associated with the ability of our model to produce a single broad O-H stretching band in the infra-red absorption spectrum. 
\end{abstract}

\maketitle

\section{Introduction}

The potential influence of quantum mechanical zero point energy and tunneling effects on the properties of liquid water has been appreciated since at least the 1970s,$^{1}$ and modern computational studies of the problem can be traced back to the pioneering path integral molecular dynamics and Monte Carlo simulations of the liquid that were performed in the mid-1980s.$^{2,3}$  However, the effects of quantum mechanical fluctuations on the properties of the liquid have still not entirely been resolved, and the importance of liquid water in a wide variety of chemical, biological, and geological contexts is continuing to motivate new research on this problem.$^{4}$

The early path integral simulations mentioned above demonstrated that quantum mechanical effects weaken the hydrogen-bonding network in liquid water and make it less structured than its classical counterpart.$^{2,3}$ More recently, the centroid molecular dynamics$^{5,6}$ (CMD) and ring polymer molecular dynamics$^{7,8}$ (RPMD) approximations have been used to study the role of quantum mechanical effects in the dynamics of the liquid.$^{9-14}$ These simulations have found that the rates of translation diffusion and orientational relaxation in ambient water increase by a factor of approximately 1.5 when quantum mechanical fluctuations are included. Similar quantum effects have also been obtained using other simulation methods, including the Feynman-Hibbs (FH) method$^{15}$ and the Feynman-Kleinert linearized path integral (FK-LPI) approach,$^{16}$ and for a variety of different water models. Overall, the general concensus that has emerged from these studies, namely that the disruption of the hydrogen-bonding network caused by quantum fluctuations leads to a less viscous liquid with faster translational diffusion and orientational relaxation, would appear to be quite well established.$^{9-16}$

There are however some reasons to believe that this may not be the whole story. First, most previous studies that have assessed the role of quantum fluctuations in liquid water have used empirical potential models that were parameterized on the basis of classical simulations. Quantum simulations of such models lead to a ``double counting'' of the quantum effects in the structure and dynamics of the liquid. This typically results in, for example, a computed diffusion coefficient for the room temperature liquid that is much larger than the experimental value, indicating that such models are unsuitable for quantum simulations. Second, many of the empirical water models that have been studied using quantum methods assume that the water monomers can be treated as rigid bodies.$^{10-12}$ Although this is convenient from the point of view of computational efficiency, it disregards the important role played by intramolecular flexibility in the liquid structure, dynamics,$^{17-19}$ and thermodynamics.$^{20,21}$ Finally, of those flexible water models that have been studied using quantum dynamics methods, the majority describe the intramolecular flexibility using simple harmonic potentials.$^{9,13,16}$ The significant anharmonicity of the O--H vibration of the water monomer, which is evident from red-shifting and broadening of the O--H stretching band in the infra-red (IR) absorption spectrum of liquid water, is thereby neglected.

In the present paper, we shall address these issues by introducing a new empirical water model that has been specifically parameterized to reproduce the liquid structure, diffusion coefficient, and IR absorption frequencies in quantum, rather than classical, simulations. Our four-site q-TIP4P/F water model is both flexible and anharmonic in the intramolecular O--H stretching vibration, and we believe that it provides a more realistic description of the effective interactions in the bulk liquid than any model that has been parameterized on the basis of classical simulations. We shall also compare and contrast our results with those of the recently developed three-site q-SPC/Fw quantum water potential,$^{13}$ and demonstrate that our model gives a better description of several properties including the melting point and the temperature dependence of the liquid density.

More importantly, we shall show that the quantum mechanical effects in our q-TIP4P/F water model are much smaller than have been observed in any previous simulation. For example, the diffusion coefficient in our quantum liquid is only around 15\% larger than it is in the corresponding classical liquid, rather than $\sim$ 50\% larger as observed in previous quantum dynamics simulations of other water models. This finding can be rationalized in terms of the competition between intermolecular and intramolecular quantum effects, and it is a direct result of using an anharmonic intramolecular potential. We find that intermolecular quantum fluctuations speed up the dynamics of the liquid by disrupting the hydrogen-bonding, in agreement with previous quantum simulations. However, quantum fluctuations in the intramolecular degrees of freedom result in a larger average water monomer dipole moment in the quantum liquid than in the classical liquid. This larger dipole moment increases the strength of the intermolecular interactions and retards the molecular motion, thereby acting in competition with the intermolecular quantum effect. This competition between intermolecular and intramolecular quantum effects has not been described before and it is the main new result of the present work.

In Section II, we briefly describe the quantum water models we have studied, and in Section III we outline our simulation procedures. In Section IV we describe the results of extensive quantum simulations of the static and dynamic properties of our q-TIP4P/F water model, paying specific attention to the comparison with both experimental results and the results of simulations with the previously-developed q-SPC/Fw model.$^{13}$ In Section V, we compare the results of classical and quantum simulations of the q-TIP4P/F model and examine the competition between intra- and intermolecular quantum effects. Section VI ends the paper with some concluding remarks.

\section{Quantum Water Models}

It is well established that when simulating inhomogeneous environments such as
the surfaces of water and ice one should use a water model that includes an explicit treatment of electronic polarization.$^{22}$ However, when simulating the bulk liquid 
one can often get away with using a simple point charge model that has been 
parameterized to capture the polarization in a mean-field sense,$^{23,24}$ 
and since this is computationally more convenient it is the approach we have adopted
in the present paper.

Of the large number of simple point charge models that have been developed 
for liquid water, the one we have chosen as the starting point for the present work
is the four-site, TIP4P/2005 classical rigid water model of Abascal and Vega.$^{25}$
Unlike most other simple point charge models, this has the advantage that it gives
a reasonable description of the ice/water phase diagram in classical Monte Carlo
simulations, correctly predicting the relative stabilities of a variety of ice
polymorphs.$^{25}$ This is an appealing feature of the model even for liquid 
simulations, because one would expect the local structures of the various ice 
polymorphs to be visited in accordance with their relative stabilities at 298 K during a simulation of the ambient liquid.

The TIP4P/2005 model is a pairwise additive intermolecular potential for the interaction between water molecules of the form$^{25}$
$$
V_{\rm inter} = \sum_{i}\sum_{j>i}\left\{4\varepsilon\left[\left({\sigma\over r_{ij}}\right)^{12}-
\left({\sigma\over r_{ij}}\right)^6\right]+\sum_{m\in i}\sum_{n\in j} {q_mq_n\over r_{mn}}
\right\},
\eqno(1)
$$
where $r_{ij}$ is the distance between the oxygen atoms and $r_{mn}$ is a distance
between partial charge sites in molecules $i$ and $j$. Two positive charges of magnitude
$q_{\rm M}/2$ are placed on the hydrogen atoms of each molecule and a negative 
charge  of $-q_{\rm M}$ is placed at a point ${\bf r}_{\rm M}$ a fraction $\gamma$ along the vector connecting the oxygen atom to the center-of-mass of the two hydrogens:
$$
{\bf r}_{\rm M} = \gamma\, {\bf r}_{\rm O}+(1-\gamma)({\bf r}_{{\rm H}_1}
+{\bf r}_{{\rm H}_2})/2. \eqno(2)
$$

In order to add intramolecular flexibility to this model, we have used a quartic 
expansion of a Morse potential to describe the stretching of the 
O--H bonds and a simple harmonic potential in the bond angle,
$$
V_{\rm intra} = \sum_i \left[V_{\rm OH}(r_{i1})+V_{\rm OH}(r_{i2})+{1\over 2}k_{\theta}(\theta_i-\theta_{\rm eq})^2\right], \eqno(3)
$$
where
$$
V_{\rm OH}(r) = D_r\left[\alpha_r^2(r-r_{\rm eq})^2-\alpha_r^3(r-r_{\rm eq})^3+{7\over 12}\alpha_r^4(r-r_{\rm eq})^4\right]. \eqno(4)
$$
Here $r_{i1}$ and $r_{i2}$ are the two O--H distances and $\theta_{i}$ is the H--O--H bond angle in water molecule $i$. The use of Eq.~(4) rather than the full Morse
potential $V_{\rm OH}(r) = D_r\left[1-e^{-\alpha_r(r-r_{\rm eq})}\right]^2$ avoids
dissociation events, which would not be described correctly using such
a simple model in any case. An anharmonic description of the O--H bonds is needed to recover the single broad absorption band that is seen in the O--H stretching region of the liquid water IR spectrum, whereas previous simulations have shown that a simple harmonic potential suffices to capture the H--O--H bending band.$^{9,13}$

In contrast to this functional form, the three-site q-SPC/Fw 
quantum water model of Voth and co-workers$^{13}$ places the negative charge on the oxygen atom ($\gamma = 1$) and uses a harmonic description of the monomer stretching and bending modes:
$$
V_{\rm OH}(r) = {1\over 2}k_r(r-r_{\rm eq})^2. \eqno(5)
$$
As has been shown previously,$^{13}$ and we shall demonstrate again below, this purely harmonic intramolecular potential does not provide a very good description of the O--H stretching band in the vibrational spectrum of liquid water. However, since the q-SPC/Fw model was also parameterized on the basis of quantum simulations, it does provide a convenient benchmark against which to compare our present model.

In total, the interaction potential in Eqs.~(1) to~(4) contains nine parameters. The intermolecular parameters are $\varepsilon$, $\sigma$, $q_{\rm M}$ and $\gamma$, all of which we have simply fixed at their values in the TIP4P/2005 potential,$^{25}$ and the intramolecular parameters are $D_{r}$, $\alpha_{r}$, $r_{\rm eq}$, $k_{\theta}$, and $\theta_{\rm eq}$. These intramolecular parameters were optimized in an extensive series of calculations to give good agreement with the experimental structure, self-diffusion coefficient, and vibrational absorption frequencies of the liquid in quantum mechanical (path integral) simulations. The final parameters of the resulting q-TIP4P/F model are listed along side those of the q-SPC/Fw model$^{13}$ in Table I.

\section{Quantum Simulation Methods}

\subsection{Path Integral Molecular Dynamics}

We have used the path integral molecular dynamics (PIMD) method$^{26}$ to
calculate a variety of structural and thermodynamic properties of the q-SPC/Fw and
q-TIP4P/F models, including radial distribution functions, dielectric constants, liquid densities, and melting points. Most of these calculations were straightforward and require little additional explanation. For example, the O--O, O--H and H--H radial distribution functions of the two liquids were obtained from 250 ps  $NVT$ path integral simulations in the presence of an Andersen thermostat.$^{27}$ Dielectric constants were obtained from much longer (10 ns) $NVT$ simulations in order to allow for complete dielectric relaxation, and liquid densities at 1 atm pressure were obtained from 10 ns $NPT$ simulations in order to fully converge the average over density fluctuations. These latter simulations were performed in the presence of both an Andersen thermostat and an isotropic Berendsen barostat.$^{28}$   

The melting point calculations were a little more complicated and do need some more explanation. These were done by performing direct coexistence simulations of the water-ice interface under atmospheric pressure.$^{29,30}$ Initial hexagonal ice configurations were generated by placing the oxygen atoms at their crystallographic sites.$^{31}$ The hydrogen atom positions were determined using the Monte Carlo procedure of Buch {\em et al.}$^{32}$ in such a way that the Bernal-Fowler rules$^{33,34}$ were satisfied and the total dipole moment of the simulation cell was exactly zero. The initial ice configuration was then equilibrated in the presence of an Andersen thermostat and an anisotropic Berendsen barostat for 50 ps,$^{28}$ before placing the secondary prismatic $(1\overline{2}10)$ face of the ice cell in contact with an equilibrated water simulation.$^{35}$ 

In total, the coexistence simulations consisted of 696 water molecules, with 360 initially in the ice phase and 336 in the liquid. The combined ice/water system was simulated for 10 ns in the presence of an Andersen thermostat and an anisotropic Berendsen barostat. Both the number density profile of oxygen atoms along the axis perpendicular to the ice/water interface and the total potential energy of the system were used as order parameters to monitor the extent of melting or freezing during the simulation.$^{29,30}$ The simulation was halted if complete melting or freezing was observed to have occurred. The melting temperature was determined to within 1 K for the q-TIP4P/F model and to within 5 K for the q-SPC/Fw model by repeating the whole procedure at different temperatures and using a bisection procedure to home in on the melting point.

\subsection{Ring Polymer Molecular Dynamics}

We have also used the ring polymer molecular dynamics (RPMD) method$^{7,8}$ to calculate several dynamical properties of the room temperature q-TIP4P/F liquid, including the self-diffusion coefficient and various orientational relaxation times. The diffusion coefficient was obtained from the time integral of the RPMD velocity autocorrelation function,
$$
D = {1\over 3}\int_0^{\infty} \tilde{c}_{\bf v\cdot v}(t)\,dt, \eqno(6)
$$
and the $l$-th order relaxation times for various axes $\eta$ of the water molecule were obtained from the time integrals of the corresponding orientational correlation functions,
$$
\tau_l^{\eta} = \int_0^{\infty} \tilde{c}_l^{\eta}(t)\,dt. \eqno(7)
$$
Both of these calculations have been described in detail in recent papers and we have used the same procedures in the present study.$^{11,36,37}$ The velocity autocorrelation function $\tilde{c}_{\bf v\cdot v}(t)$ in Eq.~(6) was calculated for 2 ps by time averaging over 100 consecutive $NVE$ RPMD trajectories of length 4 ps, with a re-sampling of the momenta from the Maxwell distribution between each trajectory. This procedure was repeated twenty times to produce an average value for and a standard error in the diffusion coefficient. The orientational correlation functions $\tilde{c}^{\eta}_l(t)$ were calculated up to 10 ps by averaging over 500 $NVE$ trajectories of length 20 ps, again with a re-sampling of the momenta between each one. Where necessary, an exponential tail was fit to the correlation function beyond 10 ps before evaluating the integral in Eq.~(7).

Voth and co-workers have recently calculated both $D$ and the same $\tau^{\eta}_l$s as us for their q-SPC/Fw water model using the centroid molecular dynamics (CMD) method.$^{13}$ This is closely related to RPMD and it gives similar results for systems like liquid water that are nearly classical. We ourselves prefer to use RPMD to calculate $D$ and $\tau^{\eta}_l$ because it gives a more accurate approximation to both $\tilde{c}_{\bf v\cdot v}(t)$ and $\tilde{c}_l^{\eta}(t)$ in the short time limit, which is the only limit in which anything has been proven rigorously about the relative accuracies of the two approximations for a general potential.$^{8}$

\subsection{Partially-Adiabatic Centroid Molecular Dynamics}

In a recent paper,$^{38}$ we have shown that IR spectra calculated using RPMD exhibit spurious peaks arising from the internal modes of the harmonic ring polymer. These peaks appear at high frequencies (above 1300 cm$^{-1}$ in liquid water at 298 K), and so do not interfere with the calculation of diffusion coefficients and orientational relaxation times, both of which are zero-frequency spectral components [see Eqs.~(6) and~(7)]. However, the spurious peaks do contaminate the O--H stretching band of the vibrational spectrum of liquid water, which is centered at around 3400 cm$^{-1}$.

A simple fix to this problem is to adjust the elements of the Parrinello-Rahman$^{26}$ mass matrix so as to shift the spurious oscillations beyond the spectral range of interest,$^{38}$ as is done in the partially-adiabatic centroid molecular dynamics (PA-CMD) method of Hone, Rossky and Voth.$^{39}$ We have therefore used this method in the present work to calculate the vibrational spectra of the q-TIP4P/F and q-SPC/Fw water models. As we have recently argued in more detail,$^{38}$ it is convenient to choose the elements of the Parrinello-Rahman mass matrix so that the internal modes of the ring polymer are shifted to a frequency of
$$
\Omega = \frac{n^{n/(n-1)}}{\beta\hbar}, \eqno(8)
$$
where $\beta=1/k_{\rm B}T$ and $n$ is the number of beads in the ring polymer ($n=32$ in the calculations reported below). This prescription gives $\Omega/2\pi c=7414$ cm$^{-1}$ at 298 K, which is sufficiently far beyond the spectral range of interest in the liquid water vibrational spectrum that increasing $\Omega$ any further does not have any significant impact on the results.$^{38}$ It also ensures that the PA-CMD simulations can be performed with the same integration time step as is used in RPMD.$^{38}$ 

The IR absorption spectra of the q-TIP4P/F and q-SPC/Fw models were calculated using the formula$^{38,40,41}$
$$
n(\omega) \alpha(\omega) = \frac{\pi \beta \omega^{2}}{3\epsilon_{0}Vc} \tilde{I}(\omega),\eqno(9)
$$
where
$$
\tilde{I}(\omega) = \frac{1}{2\pi} \int_{-\infty}^{\infty} e^{-i \omega t} \tilde{c}_{\bm{\mu}\cdot \bm{\mu}}(t)\,dt. \eqno(10)
$$
Here $n(\omega)$ is the frequency-dependent refractive index of the liquid, $\alpha(\omega)$ is a Beer-Lambert absorption coefficient, $V$ is the volume of the simulation cell, $c$ is the speed of light in a vacuum, and $\tilde{c}_{\bm{\mu}\cdot \bm{\mu}}(t)$ is a canonical (Kubo-transformed$^{42,43}$) dipole autocorrelation function. This was calculated up to 10 ps by time-averaging along 300 $NVE$ PA-CMD trajectories of length 20 ps, with a re-sampling of the momenta between each trajectory. The Fourier transform in Eq.~(10) was then performed numerically. Further details of this approach to calculating the vibrational spectrum of liquid water are given in ref.~38.

\subsection{Additional Computational Details}

Unless stated otherwise, all of our liquid simulations were performed at a temperature of 298 K and density of 0.997 g cm$^{-3}$ with 216 water molecules in a cubic simulation box. Periodic boundary conditions were applied using the minimum image convention. Short-range interactions were truncated at 9 \AA\ and Ewald summation was employed to calculate the long-range electrostatic interactions.$^{44}$ 32 ring polymer beads were used in all simulations. A ring polymer contraction scheme with a cut-off value of $\sigma = 5$ \AA\ was employed to reduce the electrostatic potential energy and force evaluations to  single Ewald sum, thereby significantly speeding up the calculations.$^{45}$ Several classical simulations were also performed for comparison by collapsing the ring polymer to a single bead ($n=1$). The ring polymer time evolution was performed analytically in the normal mode representation using a multiple time-step algorithm in which the intermolecular forces were updated every 0.5 fs and the intramolecular forces every 0.1 fs. In all simulations, the system was equilibrated for 100 ps in the presence of an Andersen thermostat$^{27}$ before the accumulation of any averages.

\section{Validation of the q-TIP4P/F model}

We shall now validate the q-TIP4P/F water model by comparing its predictions for various properties of liquid water with experimental measurements when the properties are calculated using the path integral methods described above. We shall also compare the results of this model with those of the recently-developed q-SPC/Fw quantum water model.$^{13}$ The competition between intra- and intermolecular quantum effects in the dynamics of the q-TIP4P/F model is a separate issue that we shall investigate by making a comparison with classical molecular dynamics simulations in Section~V.

\subsection{Static equilibrium properties}

Let us begin by considering the ensemble-averaged water monomer properties of the q-TIP4P/F and q-SPC/Fw water models given in Table~II. It is clear from this table that there are significant differences between the average quantum geometries of the q-TIP4P/F and q-SPC/Fw water monomers in the room temperature liquid. The bond length is greater in q-SPC/Fw by about 0.04 \AA, and the bond angle is greater by about 1.5$^{\circ}$. Despite the smaller partial charges on the q-SPC/Fw hydrogen atoms, these geometric changes result in an average monomer dipole moment in q-SPC/Fw water which is greater than that in q-TIP4P/F water by about 5\%. 

Table II also reports average values of the tetrahedral quadrupole moment$^{46}$ 
$$
Q_{T} = \frac{Q_{xx}-Q_{yy}}{2}, \eqno(11)
$$
where $Q_{xx}$ and $Q_{yy}$ are the quadrupole tensor components along the axis joining the hydrogen atoms and normal to the molecular plane. $Q_T$ gives an indication of the size of quadrupolar interactions in the liquid and has recently been shown to correlate strongly with the classical melting points of rigid-body water models.$^{47}$ The value of $\left<Q_{T}\right>$ for q-SPC/Fw is about 20\% smaller than that for q-TIP4P/F and this has significant implications for the melting points of the two models as we shall see below.

Figure~1 compares the O-O radial distribution functions (RDFs) of the two models with the results of the most recent neutron scattering experiment of Soper.$^{48}$ Both calculated O-O RDFs possess too much structure in the first peak, indicating that short-range interactions between nearest-neighbour oxygen atoms are slightly too strong; however, both models are consistent with the experiment to within its reported error bars.$^{48}$ The slightly increased structure observed in the first peak for q-TIP4P/F compared to q-SPC/Fw is as a result of its larger tetrahedral quadrupole moment (see Table II). To improve the level of agreement in this feature, a more sophisticated treatment of the short-range interactions would be necessary; recent work using Gaussian charge sites and Buckingham potentials has indeed been shown to improve matters,$^{49}$ albeit with an increase in the number of parameters defining the interaction potential. Beyond the first maximum in $g_{\rm OO}(r)$, the q-TIP4P/F model does somewhat better than the q-SPC/Fw model in reproducing the oscillations in the experimental RDF.

Our computed O--H and H--H RDFs for the two water models are compared with the neutron scattering results in Figs.~2 and~3. One sees from these figures that the experimental intramolecular O--H and H--H peak positions are correctly reproduced by the q-TIP4P/F model. This is clearly a result of our having included these RDFs in the parameterization of the model. In contrast, the q-SPC/Fw model predicts a somewhat larger O--H bond length and a larger bond angle than is seen in the experimental data, leading to intramolecular O--H and H--H peaks at radii that are slightly too large. This is consistent with the ensemble-averaged monomer geometries of the two water models given in Table~II. 

Constant-pressure PIMD simulations of both quantum water models give calculated densities at 298 K and atmospheric pressure that are in good agreement with experiment. Our calculated density for the q-SPC/Fw model is also in good agreement 
with that of an earlier path integral simulation.$^{13}$ However, as shown in Fig.~4, the density of the q-SPC/Fw water model is only really satisfactory at 298 K, the temperature at which the model was parameterized.$^{13}$ As the temperature is increased or decreased, the agreement with experiment deteriorates. In particular, our calculations indicate that the density maximum for the q-SPC/Fw model occurs at 235 K, around 40 K lower than the experimental value, and at a density which is 2\% greater than the experimental density maximum. 

In contrast, we find that the q-TIP4P/F model gives a very good description of the temperature-dependence of the liquid density. Throughout the entire range between the melting and vapourization points of liquid water at 1 atm, the difference between the q-TIP4P/F and experimental densities is less than 0.3\%. This ability of the q-TIP4P/F model to reproduce the experimental density curve is clearly inherited from the properties of the underlying rigid-body TIP4P/2005 model.$^{25}$ One reason why it is desirable to get the density curve right before simulating the ambient liquid is that it has been suggested to have some bearing on the ability of a water model to provide a good description of hydrophobic solvation.$^{50}$ 

The melting temperature of the q-TIP4P/F model obtained from our coexistence simulations of the ice-water inferface was found to be $251\pm 1$ K, around 22 K lower than the experimental value. In contrast, our coexistence simulations of the q-SPC/Fw model were found to give a melting point of $195\pm 5$ K, about 78 K lower than experiment.  The larger error bars associated with this value are a result of the slower motion at such low temperatures, which increases the time scale required to observe melting or ice lattice formation at the interface.

To put these numbers in context, classical simulations of common rigid water models have been found to give melting points which range from about 146 K for TIP3P to 272 K for TIP4P/Ice (which was specifically parameterized to reproduce this property).$^{51}$ Classical simulations have also found a strong correlation between the melting point and the tetrahedral quadrupole moment $Q_T$, with a larger quadrupole moment increasing the relative stability of the ice phase leading to a higher melting point.$^{47}$ The higher quantum melting point of the q-TIP4P/F model is therefore associated with its larger average tetrahedral quadrupole moment $\left<Q_T\right>$. It is also clearly inherited from the properties of the underlying rigid-body TIP4P/2005 model, which melts in classical simulations at 252 K.$^{25}$ Three-site models from the SPC and TIP3P families generally have smaller tetrahedral quadrupole moments and lower melting points,$^{47,51}$ and our computed quantum melting point for the q-SPC/Fw model is entirely consistent with this.

Previous classical simulations have also suggested that it is not possible to reproduce the experimental difference of 3.98 K between the temperature of maximum density and the melting point of hexagonal ice using simple point charge models.$^{52}$ The present quantum mechanical simulations of the q-TIP4P/F model give a 28 K difference between these two temperatures, which is in the middle of the 21-37 K range of differences found in classical simulations of similar models.$^{52}$ So the inclusion of quantum mechanical effects does not seem to help matters. It appears that an explicit treatment of electronic polarization will be needed to reproduce the small temperature difference between the temperature of maximum density and the melting point of water. This is consistent with the finding that the average dipole moment of the molecules in ice is significantly larger than that of the molecules in liquid water, indicating that significant charge redistribution occurs upon freezing.$^{53,54}$

Another property that is likely to benefit from an explicit treatment of electronic polarization is the dielectric constant. The final row of Table~II gives the dielectric constants obtained in the present path integral simulations of the q-TIP4P/F and q-SPC/Fw models at 298 K and 0.997 g cm$^{-3}$. These were calculated as
$$
\epsilon_{r} = 1 +{\beta\over 3\epsilon_0 V}\left(\left<\bm{\mu}\cdot\bm{\mu}\right>
-\left<\bm{\mu}\right>\cdot\left<\bm{\mu}\right>\right), \eqno(12)
$$
where $\bm{\mu}$ is the total dipole moment of the system. Our result for the q-SPC/Fw model is consistent with the value obtained by Paesani {\em et al.}$^{13}$ in an independent path integral simulation ($\epsilon_r = 86\pm 4$). Both models give only qualitatively reasonable values for the dielectric constant: q-TIP4P/F underestimates the experimental value by 22\% while q-SPC/Fw overestimates it by 15\%. These errors are not surprising given the lack of explicit electronic polarization in either model. Note also that the larger dielectric constant of the q-SPC/Fw model is consistent with its larger average monomer dipole moment. 

\subsection{Dynamical properties}

Turning now to dynamical properties, Table~III gives the results of our RPMD simulations of the diffusion coefficient and various orientational relaxation times for the q-TIP4P/F model at 298 K and 0.997 g cm$^{-3}$. The experimental diffusion coefficient of liquid H$_2$O was employed as a target during the parameterization of the model, and our calculated value of 0.221 \AA$^{2}$ ps$^{-1}$ is seen to be in good agreement with the experimental value$^{55}$ of 0.230  \AA$^{2}$ ps$^{-1}$. Perhaps more importantly, since it provides a direct experimental indication of the role of quantum mechanical effects, we obtain equally good agreement with the experimental H$_2$O/D$_2$O diffusion coefficient ratio.$^{56}$

One should however bear in mind that diffusion coefficients are quite sensitive to finite size effects,$^{11,36,57,58}$ and the simulations reported in Table~III were performed for a system of only 216 molecules. We have therefore performed some additional RPMD simulations of larger systems (containing 343 and 512 H$_2$O molecules), and extrapolated the diffusion coefficient to the limit of infinite system size using the formula$^{57,58}$ 
$$
D(\infty) = D(L)+{\xi\over 6\pi\beta\eta L}, \eqno(13)
$$
where $\eta$ is the shear viscosity, $L$ is the length of the simulation box, and $\xi$ is a numerical constant that depends on the geometry of the simulation ($\xi=2.837$ for a cubic box). This was found to give $D(\infty)=0.245\pm 0.007$ \AA$^{2}$ ps$^{-1}$ for the q-TIP4P/F model, within 7\% of the experimental result. For comparison with this, the diffusion coefficient of the q-SPC/Fw model calculated using CMD has been reported to be 0.24 \AA$^{2}$ ps$^{-1}$ for a system of 216 water molecules.$^{13}$ We would expect that this value would also increase somewhat when extrapolated to the limit of infinite system size.

The remaining rows of Table III give the orientational relaxation times for three different axes of the water molecule (the H--H axis, one of the two equivalent O--H axes, and the dipole axis) obtained from RPMD simulations of the q-TIP4P/F model, along with the CMD results for the q-SPC/Fw model from Paesani {\em et al.}$^{13}$ A previous study has shown that these orientational relaxation times are less sensitive to system size effects than the diffusion coefficient and so in this case a simulation of a system of 216 water molecules is perfectly satisfactory.$^{11}$ Comparing the calculated relaxation times with the values available from NMR and IR relaxation experiments,$^{59-62}$ we find that both the q-TIP4P/F and q-SPC/Fw models give a reasonable description of orientational relaxation in bulk liquid water in quantum simulations of the room temperature liquid.

Figure~5 compares our calculated PA-CMD dipole absorption spectra of the q-TIP4P/F and q-SPC/Fw water models with the experimental spectrum of Bertie and Lan.$^{63}$ Our spectrum for the q-SPC/Fw model is in good agreement with the fully-adiabatic CMD spectrum reported previously by Paesani {\em et al.}$^{13}$ The two calculated spectra clearly reproduce the general features of the experimental spectrum, with O-H stretching absorptions above $\sim$3000 cm$^{-1}$, a water bending band at around $\sim$1600 cm$^{-1}$, and intermolecular librational features below $\sim$1000 cm$^{-1}$. However, the peak at $\sim$200 cm$^{-1}$ is absent from both of the simulated spectra. This peak arises from the low-frequency modulation of dipole-induced dipole interactions which are clearly not present in simple point charge models.$^{64}$ The relative intensities of the intramolecular bending and stretching bands also disagree with experiment in both simulations, and this again arises from the neglect of electronic polarization. The intensity of an absorption band is proportional to the square of the derivative of the dipole moment along the corresponding vibrational coordinate, and neglecting the (quite significant) induced dipole contribution to this gives the wrong intensity.

Putting these obvious deficiencies of both models aside, the q-TIP4P/F model clearly does a better job of reproducing the experimental vibrational spectrum of liquid water than the q-SPC/Fw model. The position and width of the broad O--H stretching band in the experimental spectrum are both well reproduced by the q-TIP4P/F simulation, whereas the q-SPC/Fw model predicts distinct antisymmetric and symmetric stretching peaks that are blue-shifted relative to the experimental absorption maximum by around 80 and 190 cm$^{-1}$. The appearance of these separate peaks is a direct consequence of the use of a harmonic O--H stretching potential in the q-SPC/Fw model [see Eq.~(5)]. The position and width of the experimental intramolecular bending band are also fairly well reproduced by the q-TIP4P/F simulation, although in this case there is a slight red shift relative to the experimental absorption maximum of some 40 cm$^{-1}$. The red shift in the case of the q-SPC/Fw simulation is around 190 cm$^{-1}$. 

All of these observations are clearly consistent with the fact that the experimental absorption frequencies of liquid water were included in the data set that was used to parameterize the q-TIP4P/F model. As a further check on the reliability of the model for describing the vibrations in the liquid, we have therefore performed some additional PA-CMD simulations of heavy water, for which the model was not explicitly parameterized. The resulting q-TIP4P/F and q-SPC/Fw spectra are compared with the experimental IR spectrum$^{65}$ of D$_2$O in Fig.~6. One sees from this figure that the q-TIP4P/F model again does a better job of reproducing the positions and widths of the bands in the experimental spectrum than the q-SPC/Fw model, and indeed that the agreement between the q-TIP4P/F and experimental spectra is no worse in this case than it was for H$_2$O.

\subsection{Summary}

Combining all of the above results, it is clear that the q-TIP4P/F model does a good job of reproducing a wide variety of static and dynamic properties of liquid water in quantum mechanical (path integral) simulations. For all of the properties we have considered, the results of the q-TIP4P/F model are at least as good as those of the recently developed q-SPC/Fw quantum water model,$^{13}$ and for some properties (such as the melting point, the temperature dependence of the liquid density, and the vibrational frequencies of the liquid) they are significantly better. This is not especially surprising, as these properties were either included in the parameterization of the q-TIP4P/F model or inherited from the behavior of the underlying TIP4P/2005 potential.$^{25}$ What is surprising, however, is just how small the quantum mechanical effects in the diffusion and orientational dynamics of the liquid turn out to be when simulated with this new quantum water model, as we shall now discuss.

\section{Competing Quantum Effects}

\subsection{Computational results}

The only obvious way to estimate the effect of quantum mechanical fluctuations on the dynamics of liquid water is to compare classical (1 bead) with path integral ($n$ bead) simulations. The results of such a comparison for the diffusion coefficient of the q-TIP4P/F liquid are shown in Table~IV along with those of similar comparisons that have been made in the past for a variety of other water models.$^{9-11,14,37}$ For the q-TIP4P/F model, we find that the magnitude of the quantum effect, defined as $D_{\rm qm}/D_{\rm cl}$, is 1.11. This is significantly smaller than the effect of quantum fluctuations on the diffusion coefficient that has been found previously for any other water model, and indeed the average value of $D_{\rm qm}/D_{\rm cl}$ obtained from the earlier simulations listed in Table~IV is 1.46.

A similar result is found for orientational relaxation. Table~V compares the orientational relaxation times for three different axes of the water molecule obtained from quantum (RPMD) and classical simulations of the room temperature q-TIP4P/F liquid.  The quantum effect, defined as $\tau_{\rm cl}/\tau_{\rm qm}$, is seen to be around 1.18 for all three first order relaxation times and around 1.25 for the second order relaxation times.  These effects are again much smaller than those seen in all previous studies of orientational relaxation in liquid water,$^{10,11,13,14}$ which have typically found the quantum effect $\tau_{\rm cl}/\tau_{\rm qm}$ to be around 1.5. Clearly, then, the effect of quantum fluctuations on both the translational diffusion and the orientational relaxation of our q-TIP4P/F model is significantly smaller than has been seen previously for any other water model. But why?

\subsection{Analysis and discussion}

A clear clue as to what is going on is provided by the ensemble-averaged water monomer properties of the q-TIP4P/F model obtained in classical and quantum (PIMD) simulations of the room temperature liquid (see Table~VI). The average O-H bond length of the water molecules in the quantum liquid is greater than in the classical liquid, and the average bond angle is slightly smaller. As a result of these geometric changes, the average water molecule dipole moment in the quantum liquid is larger than that in the classical liquid by about 1.6\%. 

A similar effect has been seen previously in an {\em ab initio} PIMD study of liquid water,$^{66}$ which found the average water dipole moment to be larger than that obtained in a classical simulation by as much as 4\%. The increase in the O--H bond length in the quantum simulation is also consistent with a recent experimental investigation which has shown that the average O--H bond length in liquid H$_2$O is larger than the average O--D bond length in liquid D$_2$O.$^{67}$ The results for the q-TIP4P/F model in Table~VI are not therefore unreasonable, and indeed they arise from an intuitively rather obvious effect: zero point fluctuations in the anharmonic O--H stretching coordinate increase the average O--H bond length and give the water molecule a larger average dipole moment than it would otherwise possess.$^{66}$

The implications of all this for the magnitude of the quantum effect in the diffusion and orientational relaxation of the liquid are as follows. A larger average dipole moment leads to stronger intermolecular interactions which slow down the translational and orientational motion of the molecules in the liquid. However, there is also a competing effect: intermolecular quantum fluctuations disrupt the hydrogen bonding network leading to more rapid diffusion and orientational relaxation. For a water model in which this competing effect is the only one in operation, such as a rigid-body model or a flexible simple point charge model with a harmonic O--H stretching potential,  $D_{\rm qm}/D_{\rm cl}$ and $\tau_{\rm cl}/\tau_{\rm qm}$ are typically found to be around 1.5. But for the q-TIP4P/F model, in which both effects contribute and oppose one another, $D_{\rm qm}/D_{\rm cl}$ and $\tau^1_{\rm cl}/\tau^1_{\rm qm}$ are both reduced to around 1.15.

A simple way to verify this explanation is to turn off the intramolecular flexibility in the q-TIP4P/F model and see what happens to the ratio of the quantum and classical diffusion coefficients. This can be done by performing simulations of the liquid using the same intermolecular interactions as in the q-TIP4P/F potential, but with a fixed intramolecular geometry. We have therefore calculated both quantum (RPMD) and classical diffusion coefficients in this way, using the SHAKE$^{68}$ and RATTLE$^{69}$ algorithms to constrain the intramolecular geometry of the q-TIP4P/F model to its equilibrium value. This was found to give $D_{\rm qm}=0.440(2)$ \AA$^2$ ps$^{-1}$ and $D_{\rm cl}=0.308(3)$ \AA$^2$ ps$^{-1}$ in simulations of 216 water molecules at room temperature, and thus a quantum effect in the diffusion coefficient of $D_{\rm qm}/D_{\rm cl}=1.43$. When the competition from intramolecular quantum fluctuations is removed, we therefore recover the well-established result$^{12}$ that intermolecular quantum fluctuations speed up the dynamics of the liquid by a factor of at least 1.4.

The only remaining puzzle in Table~IV concerns the results for the TTM2.1-F model,$^{70}$ which has been found to give a quantum effect in the diffusion coefficient of $D_{\rm qm}/D_{\rm cl}=1.50$ in CMD simulations of a system of 216 water molecules.$^{14}$ This is puzzling because TTM2.1-F is a flexible and polarizable Thole-type$^{71-74}$ model with an anharmonic intramolecular potential.$^{70}$ Why, then, is there not a competition between intramolecular and intermolecular quantum effects in this model that brings $D_{\rm qm}/D_{\rm cl}$ closer to one? 

We believe that this is because the TTM2.1-F monomers dissociate to the correct products in the gas phase (uncharged radicals).$^{70,73}$ Consequently, the molecular dipole moment in this model only shows a slight increase as the O--H bond length is increased from its equilibrium value, and indeed less so than in a typical simple point charge model [see, in particular, Fig.~2(c) of ref.~75]. As a result, the energetic cost of stretching an O--H bond in the liquid is not compensated by such a favorable increase in intermolecular interactions as it is in our q-TIP4P/F model, and the O--H bonds in the TTM2.1-F liquid remain fairly close to their equilibrium values.$^{70}$ The monomers in this liquid therefore only access regions close to the minimum of the intramolecular potential energy surface where the harmonic character of the potential dominates. This is confirmed by a recent CMD simulation of the IR spectrum of the TTM2.1-F model,$^{14}$ which shows two distinct absorption bands in the O--H stretching region (a feature that is typical of harmonic models). Thus the TTM2.1-F model behaves more like the harmonic q-SPC/Fw model than our anharmonic q-TIP4P/F model in terms of both its diffusion coefficient and its IR spectrum.

A modification to the TTM2.1-F model has recently been developed with the specific aim of giving a better description of the vibrational spectra of water clusters and the bulk liquid.$^{75}$ This new TTM3-F model has a molecular dipole moment that increases more rapidly as the O--H bond is stretched, and it therefore behaves more like our q-TIP4P/F model. In particular, it gives a single broad band in the O--H stretching region of the IR spectrum in both classical and PA-CMD simulations,$^{38}$ and more recent RPMD simulations$^{76}$ have shown that it gives a quantum effect in the diffusion coefficient of $D_{\rm qm}/D_{\rm cl}\simeq 0.9$. The fact that this ratio is actually {\em less} than one implies that there must be a very strong competition indeed between intramolecular and intermolecular quantum effects in the TTM3-F model, with the intermolecular quantum effect dictating the final result.

More generally, it seems from all of the simulations we have seen that water models which give rise to two distinct peaks in the O--H stretching region of the IR absorption spectrum also give a fairly large quantum effect in the diffusion coefficient ($D_{\rm qm}/D_{\rm cl}\ge 1.4$), whereas models which give rise to a single broad absorption band in the O--H stretching region give a much smaller quantum effect ($D_{\rm qm}/D_{\rm cl}<1.2$). Since the experimental IR spectrum of liquid water shows a single broad band in the O--H stretching region, this suggests that the net quantum effect in the diffusion (and also the orientational relaxation) of the room temperature liquid may not be that large after all.

\section{Concluding Remarks}

In this paper, we have developed a new potential energy model for liquid water, q-TIP4P/F, which gives good agreement with a wide variety of static and dynamic properties of the liquid in quantum mechanical (path integral) simulations. We have also analysed the effect of quantum mechanical fluctuations on the dynamics of this model in some detail and identified two competing contributions to the quantum diffusion coefficient. Intramolecular zero point fluctuations increase the average O--H bond length and the average molecular dipole moment, leading to stronger intermolecular interactions and slower diffusion, while intermolecular quantum fluctuations disrupt the hydrogen bonding network leading to more rapid diffusion. In our q-TIP4P/F model, these two effects nearly cancel one another, leading to a comparatively small net quantum effect on the diffusion coefficient. We have argued that the same is likely to be true for other flexible water models that give a single broad absorption band in the O--H stretching region of the liquid water IR spectrum, as is seen experimentally. 

Finally, although we have focussed on the effect of quantum fluctuations on the dynamics of the liquid, we believe that the influence of these fluctuations on various static equilibrium properties is also likely to be less pronounced for the q-TIP4P/F model than for most other water models. To provide just one example of this, we have calculated the classical melting points of the q-TIP4P/F and q-SPC/Fw models by repeating the ice-water coexistence simulations described in Sec.~III.A with the ring polymers replaced by classical particles. For the q-SPC/Fw model,$^{13}$ this was found to give a classical melting point of $222\pm 2$ K, some 27 K higher than the quantum melting point. For the q-TIP4P/F model, the classical melting point was found to be $259\pm 1$ K, just 8 K above the quantum result. This is clearly a smaller quantum effect, and it is also more consistent with the experimentally-observed difference of 4 K between the melting points of H$_2$O and D$_2$O.

\begin{acknowledgements}
This work was supported by the U.S. Office of Naval Research under Contract No. N000140510460 and by the U.K. Engineering and Physical Sciences Research Council under Grant No. E01741X. We are grateful to George Fanourgakis for providing us with his TTM3-F water simulation results and to Udo Schmitt and Gunther Zechmann for helpful discussions.
\end{acknowledgements}

\begin{enumerate}
\item
F.~H.~Stillinger, \textit{Adv. Chem. Phys.} \textbf{31}, 1 (1975).
\item
R.~A.~Kuharski and P.~J.~Rossky, \textit{J. Chem. Phys.} \textbf{82}, 5164 (1985).
\item
A.~Wallqvist and B. J. Berne, \textit{Chem. Phys. Lett.} \textbf{117}, 214 (1985).
\item
F.~Paesani and G.~A.~Voth, \textit{J. Phys. Chem. B} \textbf{113}, 4017 (2009).
\item 
J.~Cao and G.~A.~Voth, \textit{J. Chem. Phys.} \textbf{100}, 5106 (1994).
\item
S.~Jang and G.~A.~Voth, \textit{J. Chem. Phys.} \textbf{111}, 2371 (1999).
\item
I.~R.~Craig and D.~E.~Manolopoulos, \textit{J. Chem. Phys.} \textbf{121}, 3368 (2004).
\item
B.~J.~Braams and D.~E.~Manolopoulos, \textit{J. Chem. Phys.}, \textbf{125}, 124105 (2006).
\item
J.~Lobaugh and G.~A.~Voth, \textit{J. Chem. Phys.} \textbf{106}, 2400 (1996).
\item
L.~Hern\'{a}ndez de la Pe\~{n}a and P.~G.~Kusalik, \textit{J. Chem. Phys.} \textbf{121}, 5992 (2004).
\item
T.~F.~{Miller III} and D.~E.~Manolopoulos, \textit{J. Chem. Phys.} \textbf{123}, 154504 (2005).
\item
L.~Hern\'{a}ndez de la Pe\~{n}a and P.~G.~Kusalik, \textit{J. Chem. Phys.} \textbf{125}, 054512 (2006).
\item
F.~Paesani, W.~Zhang, D.~A.~Case, T.~E.~{Cheatham III} and G.~A.~Voth, \textit{J. Chem. Phys.} \textbf{125}, 184507 (2006).
\item
F.~Paesani, S.~Iuchi and G.~A.~Voth, \textit{J. Chem. Phys.} \textbf{127}, 074506 (2007).
\item
B.~Guillot and Y.~Guissani, \textit{J. Chem. Phys.} \textbf{108}, 10162 (1998).
\item
J.~A.~Poulsen, G.~Nyman and P.~J.~Rossky, \textit{Proc. Natl. Acad. Sci. USA} \textbf{102}, 6709 (2005).
\item
J.-L.~Barrat and I.~R.~McDonald, \textit{Mol. Phys.} \textbf{70}, 535 (1990).
\item
A.~Wallqvist and O.~Teleman, \textit{Mol. Phys.} \textbf{74}, 515 (1991).
\item
D.~E.~Smith and A.~D.~J.~Haymet, \textit{J. Chem. Phys.} \textbf{96}, 8450 (1992).
\item
G. Rabbe and R. J. Sadus, \textit{J. Chem. Phys.} \textbf{126}, 044701 (2007).
\item
J. Lopez-Lemus, G.~A.~Chapela and J.~Alejandre, \textit{J. Chem. Phys.} \textbf{128}, 174703 (2008).
\item
T. M. Chang and L. X. Dang, \textit{Chem. Rev.} \textbf{106}, 1305 (2006).
\item
H. J. C. Berendsen, J. R. Grigera and T. P. Straatsma, \textit{J. Phys. Chem.} \textbf{91}, 6269 (1987).
\item
K. Watanabe and M. L. Klein, \textit{Chem. Phys.} \textbf{131}, 157 (1989).
\item
J.~L.~F.~Abascal and C.~Vega, \textit{J. Chem. Phys.} \textbf{123}, 234505 (2005). 
\item
M.~Parrinello and A.~Rahman, \textit{J. Chem. Phys.} \textbf{80}, 860 (1984).
\item
H.~C.~Andersen, \textit{J. Chem. Phys.} \textbf{72}, 2384 (1980).
\item
H.~J.~C.~Berendsen, J.~P.~M.~Postma, W.~F.~van Gunsteren, A.~DiNola and J.~R.~Haak, \textit{J. Chem. Phys.} \textbf{81}, 3684 (1984).
\item
T.~Bryk and A.~D.~J.~Haymet \textit{J. Chem. Phys.} \textbf{117}, 10258 (2002).
\item
R.~G.~Fernandez, J.~L.~F.~Abascal and C.~Vega, \textit{J. Chem. Phys.}, \textbf{124}, 144506 (2006).
\item
A.~Hayward and J.~R.~Reimers, \textit{J. Chem. Phys.} \textbf{106}, 1518 (1997).
\item
V.~Buch, P.~Sandler and J.~Sadlej, \textit{J. Phys. Chem. B} \textbf{102}, 8641 (1998).
\item
J.~D.~Bernal and R.~H.~Fowler, \textit{J. Chem. Phys.} \textbf{1}, 515 (1933).
\item
V.~F.~ Petrenko and R.~W. Whitworth, \textit{Physics of Ice} (Clarendon, Oxford, 1999).
\item
H.~Nada and Y.~Furukawa \textit{J. Cryst. Growth} \textbf{283}, 242 (2005).
\item
T.~F.~{Miller III} and D.~E.~Manolopoulos, \textit{J. Chem. Phys.} \textbf{122}, 184503 (2005).
\item
T.~E.~Markland and D.~E.~Manolopoulos, \textit{J. Chem. Phys.} \textbf{129}, 024105 (2008).
\item
S.~Habershon, G.~S.~Fanourgakis and D.~E.~Manolopoulos, \textit{J. Chem. Phys.} \textbf{129}, 074501 (2008).
\item
T.~D.~Hone, P.~J.~Rossky and G.~A.~Voth, \textit{J. Chem. Phys.} \textbf{124}, 154103 (2006).
\item
D.~A.~McQuarrie, \textit{Statistical Mechanics} (University Science Books, Sausalito, 2000).
\item
R. Zwanzig, \textit{Nonequilibrium Statistical Mechanics} (Oxford University Press, Oxford, 2001).
\item
R. Kubo, \textit{J. Phys. Soc. Jpn.} \textbf{12}, 570 (1957).
\item
R. Kubo, M. Toda and N. Hashitsume, \textit{Statistical Physics II: Nonequilibrium Statistical Mechanics} (Springer, New York, 1985).
\item
D.~Frenkel and B.~Smit, \textit{Understanding Molecular Simulation: From Algorithms to Applications} (Academic Press, San Diego, 2002).
\item
T.~E.~Markland and D.~E.~Manolopoulos, \textit{Chem. Phys. Lett.} \textbf{464}, 256 (2008).
\item
S.~L.~Carnie and G.~N.~Patey, \textit{Mol. Phys.} \textbf{47}, 1129 (1982).
\item
J.~L.~F.~Abascal and C.~Vega, \textit{Phys. Chem. Chem. Phys.} \textbf{9}, 2775 (2007).
\item
A.~K.~Soper, \textit{Chem. Phys.} \textbf{258}, 121 (2000).
\item
P.~Paricaud, M.~P$\Check{\text{r}}$edota, A.~A.~Chialvo and P.~T.~Cummings, \textit{J. Chem. Phys.} \textbf{122}, 244511 (2005).
\item
D.~Paschek, \textit{J. Chem. Phys.} \textbf{120}, 6674 (2004).
\item
C. Vega, E. Sanz and J. L. F. Abascal, \textit{J. Chem. Phys.} \textbf{122}, 114507 (2005).
\item
C.~Vega and J.~L.~F.~Abascal, \textit{J. Chem. Phys.} \textbf{123}, 144504 (2005).
\item
E. R. Batista, S. S. Xantheas and H. J\"onsson, \textit{J. Chem. Phys.} \textbf{109}, 4546 (1998).
\item
E. R. Batista, S. S. Xantheas and H. J\"onsson, \textit{J. Chem. Phys.} \textbf{111}, 6011 (1999).
\item
W. S. Price, H. Ide and Y. Arata, \textit{J. Phys. Chem. A} \textbf{103}, 448 (1999).
\item
W. S. Price, H. Ide and Y. Arata, \textit{J. Phys. Chem. B} \textbf{104}, 5874 (2000).
\item
B. D\"unweg and K. Kremer, \textit{J. Chem. Phys.}, \textbf{99}, 6983 (1993).
\item
I.-C. Yeh and G. Hummer, \textit{J. Phys. Chem. B}, \textbf{108}, 15873 (2004).
\item
H.~S.~Tan, I.~R.~Piletic and M.~D.~Fayer, \textit{J. Chem. Phys.} \textbf{122}, 174501 (2005).
\item
Y.~L.~A.~Rezus and H.~J.~Bakker, \textit{J. Chem. Phys.} \textbf{123}, 114502 (2005).
\item
C.~P.~Lawrence and J.~L.~Skinner, \textit{J. Chem. Phys.} \textbf{118}, 264 (2003).
\item
R.~Winkler, J.~Lindner, H.~B\"{u}rsing and P.~V\"{o}hringer, \textit{J. Chem. Phys.} \textbf{113}, 4674 (2000).
\item
J.~E.~Bertie and Z.~Lan, \textit{Appl. Spectrosc.} \textbf{50}, 1047 (1996).
\item
R. W. Impey, P. A. Madden and I. R. McDonald, \textit{Mol. Phys.} \textbf{46}, 513 (1982).
\item
J. E. Bertie, M. K. Ahmed and H. H. Eysel, \textit{J. Phys. Chem.} \textbf{93}, 2210 (1989).
\item
B.~Chen, I.~Ivanov, M.~L.~Klein and M.~Parrinello, \textit{Phys. Rev. Lett.} \textbf{91}, 215503 (2003).
\item
A.~K.~Soper and C.~J.~Benmore, \textit{Phys. Rev. Lett.} \textbf{101}, 065502 (2008).
\item
J.-P.~Ryckaert and G.~Ciccotti and H.~J.~C.~Berendsen, \textit{J. Comput. Phys.} \textbf{23}, 327 (1977).
\item
H.~C.~Andersen, \textit{J. Comput. Phys.} \textbf{52}, 24 (1983).
\item
G. S. Fanourgakis and S. S. Xantheas, \textit{J. Phys. Chem. A} \textbf{110}, 4100 (2006).
\item
B.~T.~Thole, \textit{Chem. Phys.} \textbf{59}, 341 (1981).
\item
C.~J.~Burnham, J.~C.~Li, S.~S.~Xantheas and M.~Leslie, \textit{J.~Chem.~Phys.} \textbf{110}, 4566 (1999).
\item
C.~J.~Burnham and S.~S.~Xantheas, \textit{J.~Chem.~Phys.} \textbf{116}, 5115 (2002).
\item
C.~J.~Burnham, G.~F.~Reiter, J.~Mayers, T.~Abdul-Redah, H.~Reichert and H.~Dosch,
\textit{Phys. Chem. Chem. Phys.} \textbf{8}, 3966 (2006).
\item
G.~S.~Fanourgakis and S.~S.~Xantheas, \textit{J. Chem. Phys.} \textbf{128}, 074506 (2008).
\item
G.~S.~Fanourgakis, private communication (2008).
\item
A. Saul and W. Wagner, \textit{J. Phys. Chem. Ref. Data} \textbf{18}, 1537 (1989).
\item
R. C. Weast, ed., \textit{Handbook of Chemistry and Physics} (58$^{\rm th}$ Edition, CRC, Cleveland, 1977).
\end{enumerate}

%
%
\newpage
\begin{table}[htp]
\setlength{\extrarowheight}{5pt}
\setlength{\tabcolsep}{15pt}
\begin{center}
\begin{tabular}{ccc}
\hline \hline
                                         & q-TIP4P/F    &   q-SPC/Fw \\
\hline
$\varepsilon$ / kcal mol$^{-1}$           & 0.1852        & 0.1554 \\
$\sigma$ / \AA                             & 3.1589         & 3.1655 \\
$q_{\rm M}$ / $|e|$                          & 1.1128        & 0.84 \\
$\gamma$                                       & 0.73612       & 1.00 \\
$D_{r}$ / kcal mol$^{-1}$              & 116.09         & -\\
$\alpha_{r}$ / \AA$^{-1}$         & 2.287           & - \\
$k_{r}$ / kcal mol$^{-1}$\AA$^{-2}$    & -          & 1059.162  \\
$r_{\rm eq}$ / \AA                             & 0.9419         & 1.0000 \\
$k_{\theta}$  /  kcal mol$^{-1}$ rad$^{-2}$         & 87.85 & 75.90  \\
$\theta_{\rm eq}$ / degrees               & 107.4          & 112.0 \\ 
\hline \hline
\end{tabular}
\caption{Parameters in the q-TIP4P/F and q-SPC/Fw (ref.~13) quantum water models.}
\end{center}
\end{table}
\clearpage
%
%
\begin{table}[htp]
\setlength{\extrarowheight}{5pt}
\setlength{\tabcolsep}{15pt}
\begin{center}
\begin{tabular}{cccc} \hline\hline
 & q-TIP4P/F & q-SPC/Fw & Experiment   \\
\hline
$\langle r_{\rm OH}\rangle $ / \AA & 0.978(1) & 1.019(1) & 0.97$^a$ \\
$\langle\theta_{\rm HOH}\rangle$ /  degrees & 104.7(1) & 106.2(1) & 105.1$^a$ \\
$\langle\mu\rangle$ / D & 2.348(1) & 2.465(1) & - \\
$\langle Q_{T}\rangle$ / D \AA & 2.403(1) & 2.009(1) & - \\
$\rho_{298}$ / g cm$^{-3}$ & 0.998(2) & 1.002(2)  & 0.997$^{b}$ \\
$\rho_{\rm TMD}$ / g cm$^{-3}$ &  1.001(2) & 1.020(2) & 1.000$^{b}$  \\
$T_{\rm TMD}$  / K & 279(2) & 235(2) & 277.13$^{b}$  \\
$T_{\rm melt}$ / K & 251(1) & 195(5) & 273.15$^b$ \\ 
$\epsilon_r$ & 60(3) & 90(3) & 78.4$^c$ \\
\hline\hline
\end{tabular}
\caption{Static equilibrium properties of the q-TIP4P/F and q-SPC/Fw quantum water models obtained from PIMD simulations. The standard errors in the final digits are given in parentheses. $\langle \mu\rangle$ and $\langle Q_{T} \rangle$ are the ensemble-averaged values of the molecular dipole and tetrahedral quadrupole moments, $\rho_{\rm 298}$ is the liquid density at 298 K and 1 atm pressure, $\rho_{\rm TMD}$ is the density at the temperature of maximum density, $T_{\rm melt}$ is the melting point, and $\epsilon_r$ is the dielectric constant. The experimental values are from (a) ref.~48, (b) ref.~77, (c) ref.~78.}
\label{table:2}
\end{center}
\end{table}
\clearpage
%
%
\begin{table}[htp]
\setlength{\extrarowheight}{5pt}
\setlength{\tabcolsep}{15pt}
\begin{center}
\begin{tabular}{cccc} \hline\hline
    & q-TIP4P/F &  q-SPC/Fw & Experiment   \\
\hline
$D_{{\rm H}_2{\rm O}}$ / \AA$^{2}$ ps$^{-1}$ & 0.221(1) & 0.24(1) &    0.230$^a$ \\
$D_{{\rm D}_2{\rm O}}$ / \AA$^{2}$ ps$^{-1}$ & 0.172(1) & - &   0.177$^a$ \\
$D_{{\rm H}_2{\rm O}}$/$D_{{\rm D}_2{\rm O}}$ & 1.28 & - &   1.30$^a$ \\ 
$\tau_{1}^{\rm HH}$ / ps & 5.40(4) & 3.8(1)  &  - \\
$\tau_{1}^{\rm OH}$ / ps & 5.10(4) & 4.1(1)  &  - \\
$\tau_{1}^{\mu}$ / ps & 4.64(3) & 4.7(1)  &  - \\
$\tau_{2}^{\rm HH}$ / ps & 2.22(2) & 1.85(5) &  1.6-2.5$^b$ \\
$\tau_{2}^{\rm OH}$ / ps & 1.90(2) & 1.70(5) &  1.95$^b$ \\
$\tau_{2}^{\mu}$ / ps & 1.52(1) & 1.60(5) &  1.90$^b$ \\
\hline\hline
\end{tabular}
\caption{Dynamical properties of liquid water at 298 K and 0.997 g cm$^{-3}$ (and heavy water at 1.107 g cm$^{-3}$) obtained from RPMD simulations of the q-TIP4P/F model. $D$ is the diffusion coefficient and $\tau_{l}^{\eta}$ is the $l$-th order orientational relaxation time for molecular axis $\eta$. The CMD results for the q-SPC/Fw model from ref.~13 are provided for comparison. The experimental values are from (a) refs.~55 and 56, (b) refs.~59-62.}
\label{table:3}
\end{center}
\end{table}
\clearpage
%
%
\begin{table}[htp]
\setlength{\extrarowheight}{5pt}
\setlength{\tabcolsep}{10pt}
\begin{center}
\begin{tabular}{cccccc}
\hline
\hline
Potential  & $N$  & Method & $D_{\rm cl}$ / \AA$^{2}$ ps$^{-1}$ & $D_{\rm qm}$ / \AA$^{2}$ ps$^{-1}$ & $D_{\rm qm} / D_{\rm cl}$ \\
\hline
q-TIP4P/F & 216   & RPMD     &  0.199(2)    & 0.221(1)  & 1.11 \\
SPC/E  & 216 & RPMD$^a$ & 0.242(3)  & 0.343(2) & 1.42  \\
SPC/F  & 216 & RPMD$^b$ & 0.279(2)   & 0.400(3)  & 1.43 \\ 
SPC/F  & 125 & CMD$^c$   & 0.30(2) & 0.42      & 1.40 \\
TTM2.1-F  & 216 & CMD$^d$ & 0.150(5) & 0.225(5) & 1.50 \\
TIP4P        & 256 & CMD$^e$ & 0.358 & 0.548 & 1.53 \\
\hline
\hline
\end{tabular}
 \caption{Classical versus quantum diffusion coefficients of several liquid water models, including the q-TIP4P/F model developed in this work. The second column gives the number of water molecules used in each simulation, and the third indicates the approximate quantum dynamical method employed; further details can be found in (a) ref.~11, (b) ref.~37, (c) ref.~9, (d) ref.~14 and (e) ref.~10. Where given, the number in parentheses indicates the standard error in the final digit. The last column gives the magnitude of the quantum effect, defined as $D_{\rm qm} / D_{\rm cl}$. The average quantum effect from the earlier studies in the table is 1.46.}
\label{table:5}
\end{center}
\end{table}
\clearpage
%
%
\begin{table}[htp]
\setlength{\extrarowheight}{5pt}
\setlength{\tabcolsep}{10pt}
\begin{center}
\begin{tabular}{cccc}
\hline
\hline
           & $\tau_{\rm cl}$ / ps  &  $\tau_{\rm qm}$ / ps & $\tau_{\rm cl} / \tau_{\rm qm}$  \\
\hline
$\tau_{1}^{\rm HH}$    & 6.3(1)  & 5.40(4) & 1.17  \\
$\tau_{1}^{\rm OH}$    & 6.0(1)  &  5.10(4) & 1.18 \\
$\tau_{1}^{\mu}$         &  5.48(7) & 4.64(3) &  1.18 \\
$\tau_{2}^{\rm HH}$    & 2.72(4) & 2.22(2) & 1.23  \\
$\tau_{2}^{\rm OH}$    & 2.38(4)  &  1.90(2) & 1.25 \\
$\tau_{2}^{\mu}$         & 1.92(3)  & 1.52(1) & 1.26 \\
\hline
\hline
\end{tabular}
 \caption{Classical versus quantum (RPMD) orientational relaxation times for the q-TIP4P/F water model at 298 K and 0.997 g cm$^{-3}$. The numbers in parentheses are the standard errors in the final digits. The final column gives the ratio of the classical and quantum relaxation times as a measure of the magnitude of the quantum effect in orientational relaxation.}
\label{table:6}
\end{center}
\end{table}
\clearpage
%
%
\begin{table}[htp]
\setlength{\extrarowheight}{5pt}
\setlength{\tabcolsep}{10pt}
\begin{center}
\begin{tabular}{ccc}
\hline
\hline
           & Classical   &  PIMD  \\
\hline
$\left< r_{\rm OH} \right>$ / \AA      &  0.963(1)      &   0.978(1)   \\
$\left< \theta_{\rm HOH} \right>$ / degrees   &   104.8(1)      & 104.7(1)   \\
$\left< \mu \right>$ / D     &  2.311(1)  & 2.348(1)    \\
\hline
\hline
\end{tabular}
 \caption{Water monomer properties for the q-TIP4P/F model in classical and quantum (PIMD) simulations. Calculations were performed at a temperature of 298 K and density of 0.997 g cm$^{-3}$. The standard errors in the final digits are given in parentheses.}
\label{table:4}
\end{center}
\end{table}
\clearpage

\newpage
\newlength{\figwidth}
\setlength{\figwidth}{0.75\columnwidth}
%
%
\begin{figure}[h]
\centering
\resizebox{\figwidth}{!} {\includegraphics{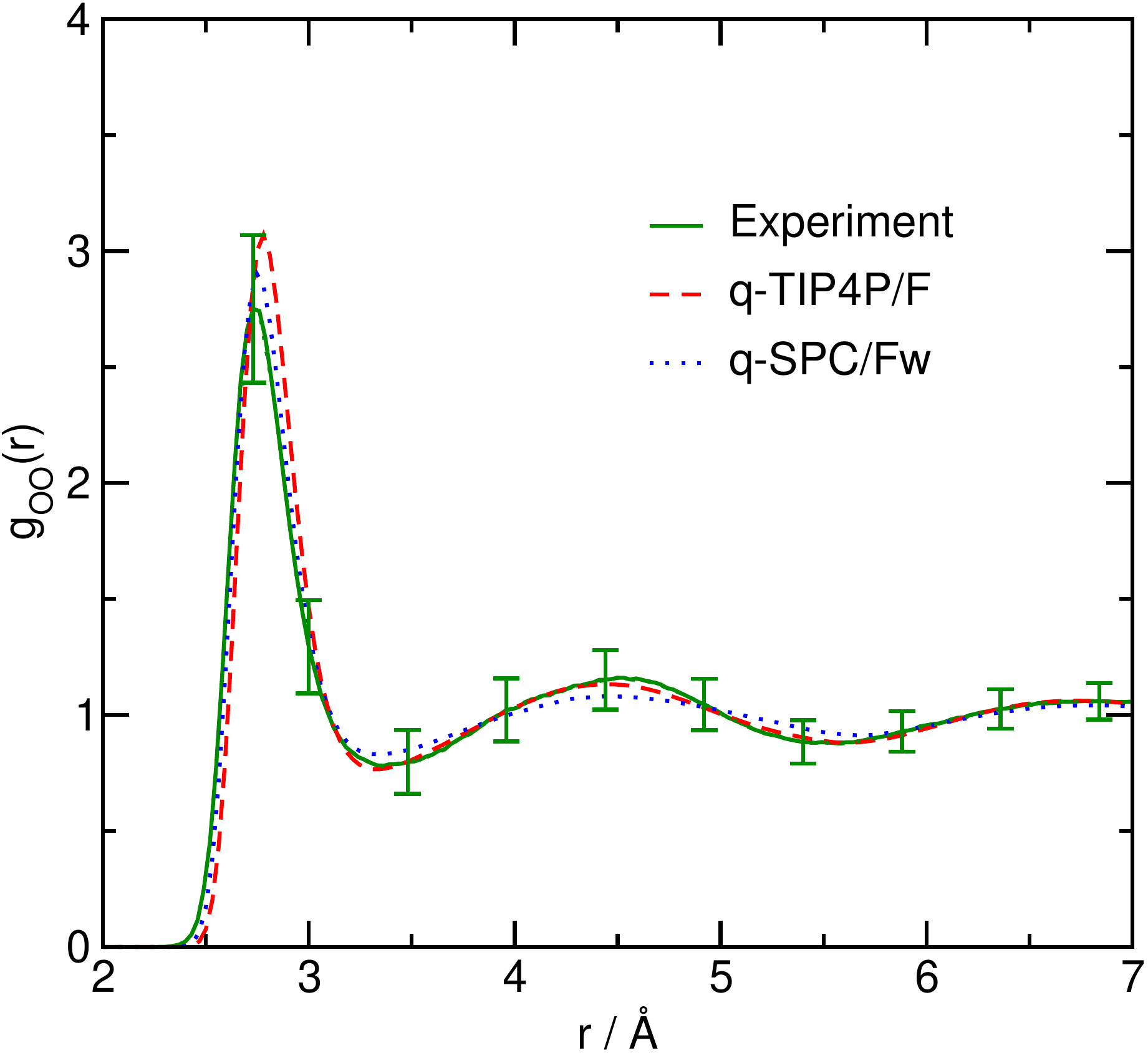}}
\caption{\label{fig:1}(Color online.) Oxygen-oxygen radial distribution functions of the q-TIP4P/F and q-SPC/Fw quantum water models obtained from PIMD simulations at 298 K and 0.997 g cm$^{-3}$. The experimental radial distribution function from ref.~48 is shown for comparison (along with its associated error bars).}
 \end{figure}
%
%
\begin{figure}[h]
\centering
\resizebox{\figwidth}{!} {\includegraphics{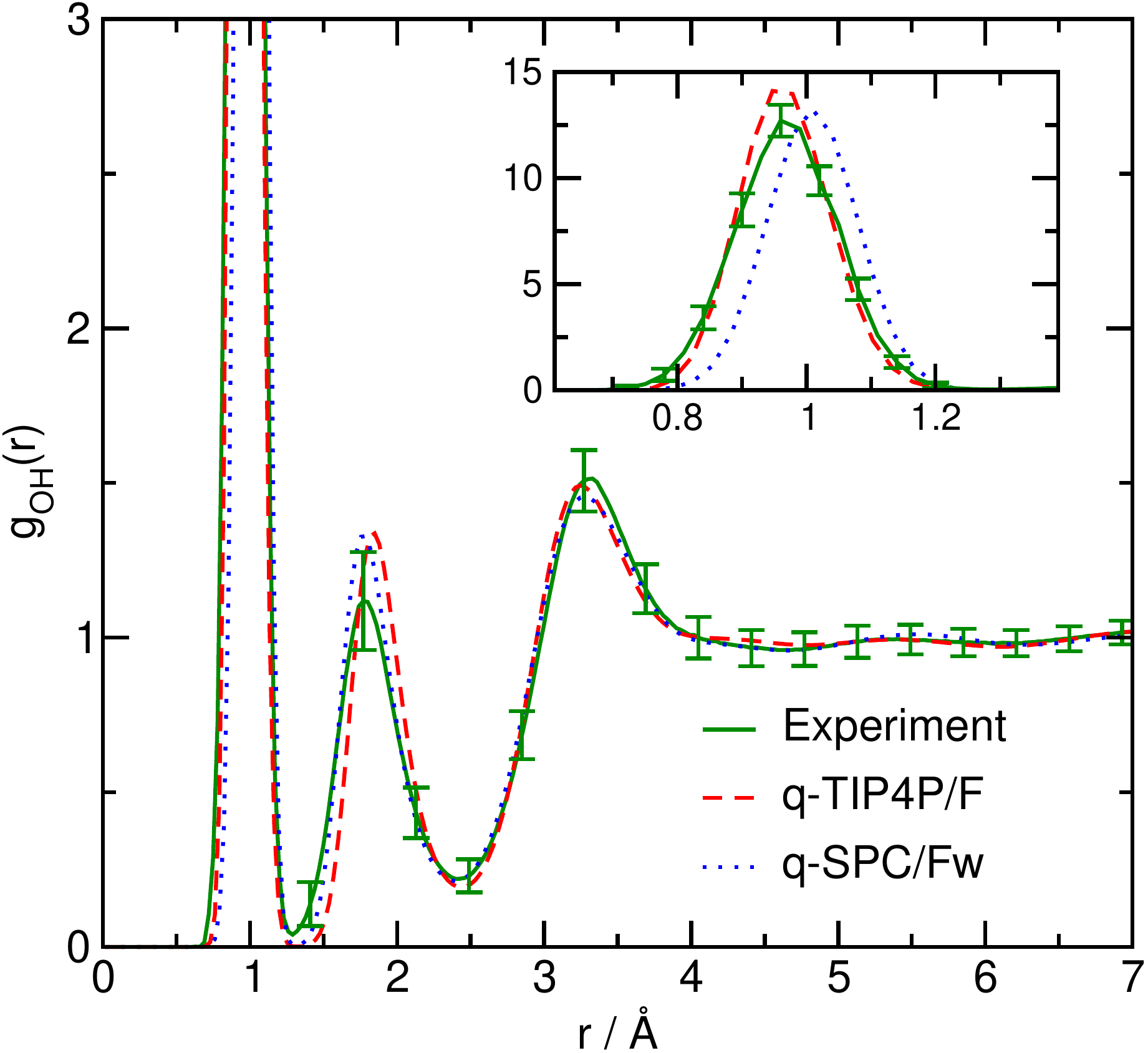}}
\caption{\label{fig:2}(Color online.) Oxygen-hydrogen radial distribution functions of the q-TIP4P/F and q-SPC/Fw quantum water models obtained from PIMD simulations at 298 K and 0.997 g cm$^{-3}$. The experimental radial distribution function from ref.~48 is shown for comparison (along with its associated error bars). The inset shows the intramolecular O--H peak at distances close to 1 \AA.}
 \end{figure}
%
%
\begin{figure}[h]
\centering
\resizebox{\figwidth}{!} {\includegraphics{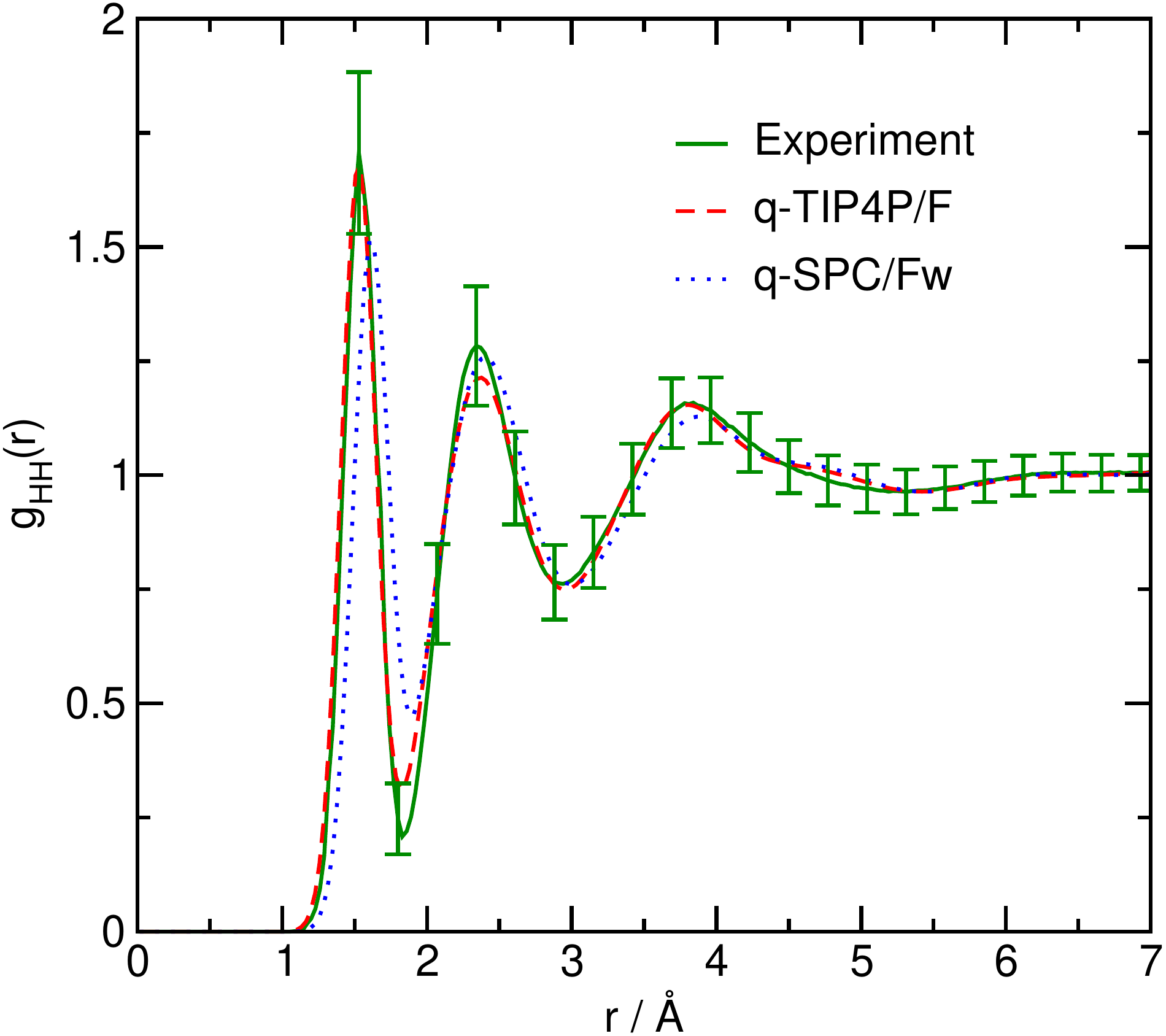}}
\caption{\label{fig:3}(Color online.) Hydrogen-hydrogen radial distribution functions of the q-TIP4P/F and q-SPC/Fw quantum water models obtained from PIMD simulations at 298 K and 0.997 g cm$^{-3}$. The experimental radial distribution function from ref.~48 is shown for comparison (along with its associated error bars).}
 \end{figure}
%
%
\begin{figure}[h]
\centering
\resizebox{\figwidth}{!} {\includegraphics{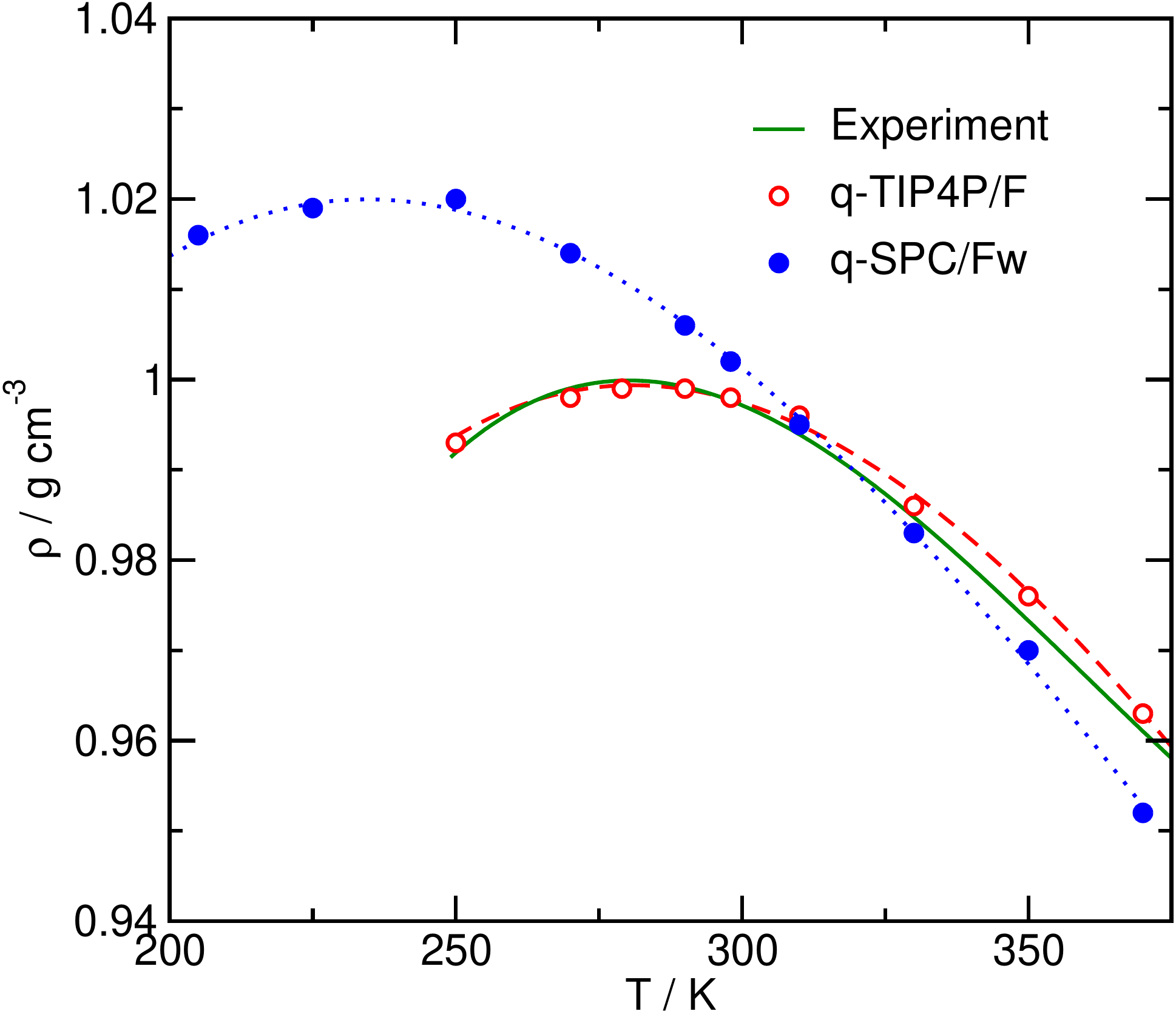}}
\caption{\label{fig:4}(Color online.) Densities at 1 atm pressure of the q-TIP4P/F and q-SPC/Fw water models obtained from PIMD simulations. The experimental density curve from ref.~77 is shown for comparison. The red and blue curves through the calculated points are simply guides for the eye.}
 \end{figure}
%
%
\begin{figure}[h]
\centering
\resizebox{\figwidth}{!} {\includegraphics{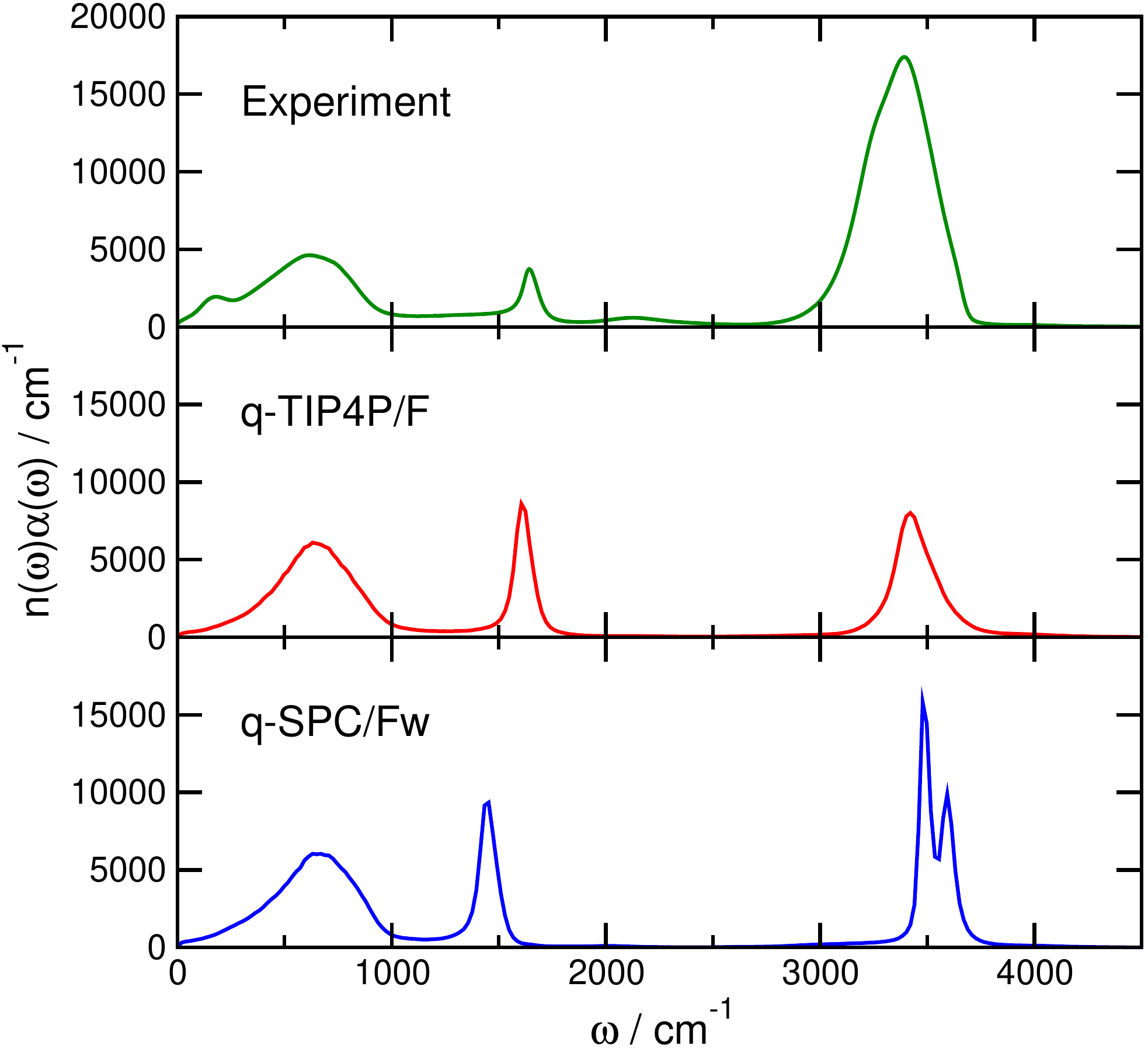}}
\caption{\label{fig:5}IR absorption spectra of the q-TIP4P/F and q-SPC/Fw water models obtained from PA-CMD simulations at 298 K and 0.997 g cm$^{-3}$. The upper panel shows the experimental spectrum from ref.~63.}
 \end{figure}
%
%
\begin{figure}[h]
\centering
\resizebox{\figwidth}{!} {\includegraphics{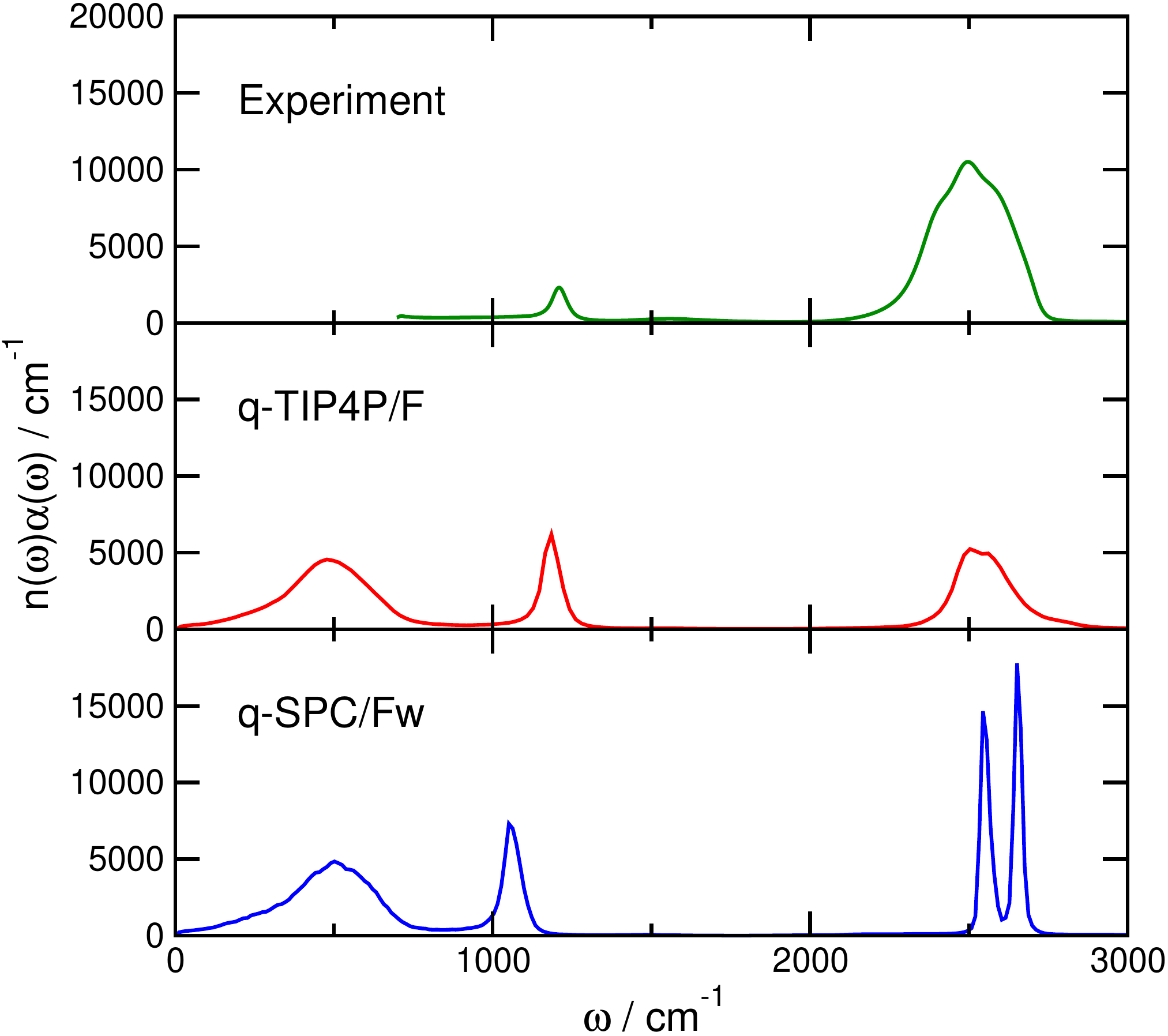}}
\caption{\label{fig:6}As in Fig.~5, but for D$_2$O at a temperature of 298 K and a density of 1.107 g cm$^{-3}$. In this case the experimental spectrum is only available above 700 cm$^{-1}$ (ref.~65).}
\end{figure}

\end{document}